# Terahertz Waveguiding in Silicon-Core Fibers

Derek A. Bas,[1] Scott K. Cushing,[1] John Ballato,[2] and Alan D. Bristow[1,*]

[1]*Department of Physics, West Virginia University, Morgantown, WV, 26506, USA*
[2]*Center for Optical Materials Science and Engineering Technologies (COMSET), School of Materials Science and Engineering, Clemson University, Clemson, SC 29631, USA*
[*]*alan.bristow@mail.wvu.edu*

**Abstract:** We propose the use of a silicon-core optical fiber for terahertz (THz) waveguide applications. Finite-difference time-domain simulations have been performed based on a cylindrical waveguide with a silicon core and silica cladding. High-resistivity silicon has a flat dispersion over a 0.1 – 3 THz range, making it viable for propagation of tunable narrowband CW THz and possibly broadband picosecond pulses of THz radiation. Simulations show the propagation dynamics and the integrated intensity, from which transverse mode profiles and absorption lengths are extracted. It is found that for 140 – 250 μm core diameters the mode is primarily confined to the core, such that the overall absorbance is only slightly less than in bulk polycrystalline silicon.

## 1. Introduction

The terahertz (THz) frequency range is useful for various technological and scientific applications; see [1,2] and references therein. THz radiation lies between the microwave and infrared (IR) regions of the electromagnetic spectrum and so THz-based technologies share many of the same potential applications as its spectral neighbors. However, THz components are less developed and so many hurdles exist that must be overcome in order to develop suitable device architectures. For example, and most simply, high-quality transmission fibers are needed to confine and direct THz radiation in a manner that is robust and enabling of compact practical devices. Various schemes have been investigated for THz waveguiding using either dielectric [3-6] and metallic [7] materials. Dielectric fibers provide advantages in length and flexibility, extend low-frequency waveguiding capabilities, and have much lower power absorption than metal waveguides [4]. Existing schemes are typically very efficient for laboratory use in the specific applications for which they are designed, with attenuation coefficients as low as about 0.01cm$^{-1}$ [3,5]. However, designing a multipurpose, commercially-manufactured THz waveguide would be an important advancement of the field. Flexibility and available wavelength ranges become important considerations and for applications such as endoscopy [8] it is important that the waveguide's core, and hence the propagating mode, be protected by a cladding for deployment.

It has been predicted that silicon would be an optimal material for THz applications. Factors that support the claim are its relatively constant refractive index and low absorption over the THz frequency range [9], as well as its high thermal conductivity, high optical damage threshold, and low-loss transmission in the infrared frequency range [10]. Recently, Ballato, *et al.*, have manufactured various silicon-core fibers using a scalable molten-core technique, which shows promising results for future applications [10-12]. Alternative manufacturing techniques include chemical vapor deposition [13] and high pressure melt infiltration [14]. The molten-core method is particularly appropriate for THz applications, because it provides for large core (> 50 μm) fiber fabrication (potentially single-mode at THz frequencies) with length scalability and potentially high-speed manufacturing.

Silicon-core optical fibers have been shown to guide near- and mid-wave infrared light, increase nonlinear-optical properties [15] and act as Faraday isolators [16]. The purpose of this paper is to extend the range of potential applications for silicon optical fiber to THz waveguides. The propagation of pulsed THz radiation is simulated to capture the propagation effects over a wide tuning range for either narrow band continuous-wave (CW) or broadband pulse transmission. Proof-of-concept studies are being performed and the observed THz transmission spectra and properties will be discussed in detail elsewhere.

## 2. Mode Analysis

Similarly to conventional optical fibers, semiconductor-core waveguides for THz applications operate on the (classical) principal of total-internal reflection. The radiation modes are confined and guided by the higher index silicon core ($n_{core}$ = 3.418 at ~1 THz) [9], which is clad by silica ($n_{clad}$ = 1.954 at ~1 THz) [9,17]. Optical fibers are characterized by a normalized frequency parameter $V = \pi(NA)d/\lambda$, where the core diameter, $d$, is taken for the purposes of this work to be 140, 170, 200, and 250μm. The wavelength $\lambda$, is defined as $c/\nu n_{core}$, with $c$ being the vacuum light speed and the frequency range $\nu = 0.5 - 3$ THz, which correspond to core wavelengths of ~29.2 μm – ~175 μm. The numerical aperture for an optical fiber is given by $NA \approx \sin\theta_c = \sqrt{n_{core}^2 - n_{clad}^2}/n_0$, where $\theta_c$ is the half acceptance angle and $n_0$ for the external medium, which is usually air. The core-cladding refractive index contrast is >1, which leads to a very high $NA$ compared to conventional optical fibers, where the refractive index contrast is typically ~$10^{-3}$. For practical considerations, such as for single-mode operation, to eliminate intermodal dispersion, and to increase output efficiency, it is preferable to minimize $V$, hence $NA$. This can be accomplished by setting $n_0 = 3.418$ to simulate end-butt coupling to a silicon chip or hemispherical lens.

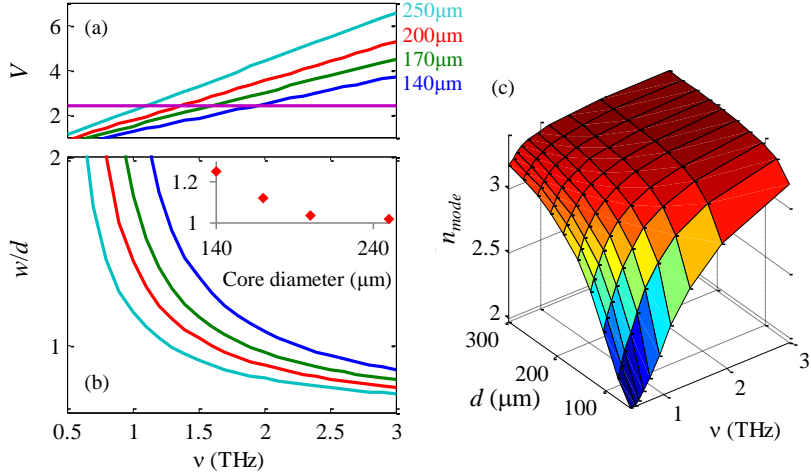

Fig. 1 (a) Normalized frequency parameter $V$ for fibers with $d = 140, 170, 200$, and 250 μm (blue, green, red, and cyan). The horizontal line is at V = 2.405. (b) The normalized $1/e^2$-diameter of fundamental $HE_{11}$ mode. Inset shows simulated Gaussian pulse values of $w/d$ versus core thickness. (c) Modal refractive index THz frequency and core diameters.

$V$ is graphed as a function of frequency in Fig. 1(a), showing a linear increase in the V with frequency and a steeper slope with increasing core diameter. A fiber is single mode if its normalized frequency is less than the first root of the $J_0$ Bessel function, i.e. $V < 2.405$, which is indicated as the horizontal line on the plot. Based on the normalized frequency calculations, it was determined that for $d < 140$ μm, wave propagation will have too much spatial overlap into the cladding, leading to higher attenuation. On the other hand, for $d > 140$ μm the wavelength range for single-mode operation diminishes.

Using commercially available OptiMode software, the fundamental $HE_{11}$ guided mode profile was computed. The $1/e^2$-diameter of the mode profile was found in terms of the normalized frequency parameter and core diameter [18] as

$$w = d\left(0.65 + 1.619V^{-1.5} + 2.879V^{-6}\right). \quad (1)$$

The normalized mode profile $w/d$ is plot as a function of frequency in Fig. 1(b). The $w/d$ parameter increases geometrically with decreasing frequency, but more slowly for wider core diameters. For high frequencies, $\nu > 1.5$ THz, $w/d$ ranges from 0.75 –1.4. Values of $w/d \gg 1$ leads to overlap of the mode in the cladding, which for THz radiation will result in absorption in the silica [9,17]. This issue is limited for wider cores, but even the 250 μm core shows overlap into the silica.

The modal index $n_{mode}$ was determined from the spatial extent of the mode profile; see Fig. 1(c). It is found that for high frequencies and large core diameters $n_{mode}$ approaches the value of bulk silicon, $n_{core}$. At low frequencies and especially in smaller core fibers there is increased mode overlap with the cladding and the mode index approaches $n_{clad}$, confirming the result of the normalized $1/e^2$-diameter of the mode profile.

Fresnel losses occur at the interfaces between external Si (chip or hemisphere) and fiber, due to mismatch between the $n_{mode}$ and $n_0$. For example, 0.5-THz radiation coupling to the 140-μm fiber has a reflection of $I_R/I_0 = (n_{si}^2 - n_{mode}^2)/(n_{si}^2 + n_{mode}^2) \approx 0.03$. Reflection decreases as the core diameter and frequency increase, because $n_{mode} \to n_0$.

## 2. Finite-Difference Time-Domain Simulations

Finite-difference time-domain (FDTD) simulations also were performed on fibers with core diameters of $d = 140, 170, 200,$ and 250μm, and cladding outer diameters of 1.6 μm using open-source Meep software [19]. Due to the radial symmetry of the optical fiber geometry two-dimensional simulations are sufficient to determine propagation trends. The simulation cell has a grid size of 5000 μm × 3000 μm, surrounded by a 100-μm border to absorb all reflections into the simulation region. The spatial and temporal resolution is set to 6.25 μm and 41.7 fs respectively.

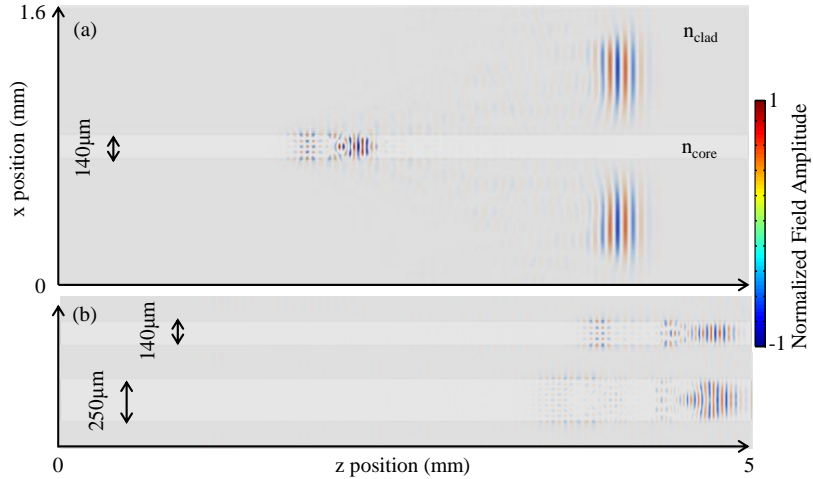

Fig. 2 (a) Spatial pattern of intensity at a single instant in time for a pulse travelling through a 140μm core fiber. Larger darkened areas are a visual aid to show the cladding and should not be confused with a nonzero intensity (Media 1; 6.6 MB). (b) Comparison of main pulse and dispersive low-frequency tail for smallest and largest core diameters (Media 2; 6.6 MB).

The core is centrally located inside the cell with dimensions of 4000 μm × $d$ and symmetric cladding has dimensions of 4000 μm × 1600 μm. A bulk Si region is placed at the entrance and exit regions of the fiber region to mimic either end-butt coupled focusing hemispheres or planar chip. The dielectric constants are determined from fitting the literature results [9] with a damped oscillator model to determine the dispersion relation

$$\varepsilon(\omega) = \varepsilon(\infty) + \frac{A\omega_0^2}{\omega_0^2 - \omega^2 - i\omega\gamma}, \quad (2)$$

where the high-frequency dielectric constant is $\varepsilon(\infty) = 11.67895$, the oscillator amplitude is $A = 0.00248$, resonance frequency is $\omega_0 = 2.35368$ THz and damping rate is $\gamma = 17.1509$ THz for the Si core. Similarly, $\varepsilon(\infty) = 3.64407$, $A = 0.16585$, $\omega_0 = 5.47686$ THz and $\gamma = 2.96159$ THz for the $SiO_2$ cladding. These units are scaled to dimensionless parameters to run in the FDTD code.

The source is described as the discrete time derivative of a Gaussian [19] waveform as

$$E(t) = (-i\omega)^{-1} \frac{\partial}{\partial t} \exp(-i\omega t - (t-t_0)^2 / 2w^2), \qquad (3)$$

which corresponds to a plane-wave source with a Gaussian envelope determined from the frequency-domain properties of a desired pulse with center frequency of 1.755 THz and the full-width at half-maximum (FWHM) of 2.925 THz. These parameters give a pulse width of $\tau_p \sim 342$ fs. This pulse shape approximately reproduces those found by optical rectification; for example see [20]. The input extends across the full input (left) side of the cell, such that the radiation couples to both the core and cladding. This assumes that the mode is nearly equivalent to being at the beam waist if it were a focused Gaussian mode, and also provides a faster travelling reference pulse in the cladding for comparison to the core region of the fiber. From the source radiation, the algorithm incrementally evolves over finite time intervals using Maxwell's equations [18], propagating in the $+z$ direction. The FDTD simulation output electric field snapshots separated by about 300 fs in time.

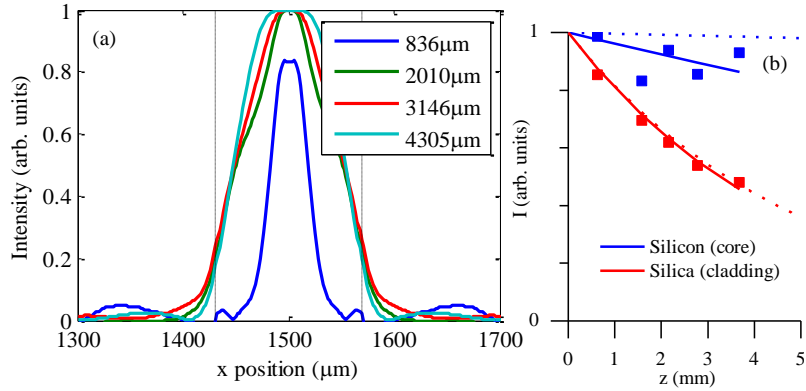

Fig. 3. (a) Evolving shape of the pulse as the peak electric field reaches z-distances given in the legend. Vertical dashed lines indicate the core-cladding boundary. (b) Attenuation of core (blue) and cladding (red) modes. The blue squares represent the flux through the core, and the red squares represent the flux through the entire fiber. Blue and red solid lines are exponential fits to the data points. Blue and red dotted lines are Beer's law attenuation curves for pure silicon and silica [9,16], respectively.

Figure 2(a) shows the full-cell intensity snapshot at a single time step (17.67 ps after the wave enters the fiber) for $d = 140$ μm. Regardless of the core diameter, the bulk of the pulse propagating in the core has a speed of $0.29c$, which is consistent with $n_{core} = 3.45$. Lower frequency components clearly lag the main pulse at a slower velocity, which is consistent with the determination of $V$. In comparison, the cladding propagation speed is $\sim 0.6c$, due to the lower index. Hence, core waveguiding can be separated from cladding propagation by taking into account the refractive index contrast between the cladding and the fundamental waveguiding mode. For example, in a 1cm long large diameter fiber (for $n_{mode} \approx n_{core}$), the pulses that travel in the fundamental mode and in the cladding should be separated in time by $z/c(n_{mode} - n_{clad}) = 48.8$ ps. This temporal separation is an advantage for time-domain THz spectroscopy, which could easily identify each component.

A side-by-side comparison of the pulse near the exit of the smallest and largest core diameters at the same time step is shown in Fig. 2(b). A clear tail is present for the lower frequency components in both fibers. In the 140 μm fiber the tail is dominated by poorer mode confinement, while in the 250 μm fiber intermodal dispersion is more prominent, although both effects are present in each fiber.

The $1/e^2$ diameter of the mode is extracted by examining the transverse position $x$ extracted from the intensity profile for a given time slice, wherein the $z$ position is selected to be the peak THz intensity. Figure 3(a) shows the extracted intensity profiles for various positions $z$ = 836, 2010, 3146 and 4305 μm. Shortly after the pulse enters the core, its profile is very narrow and well-confined to the core, but after travelling approximately 4 mm it fills out the core and the shape becomes relatively stable.

When the profile is stable, a $w/d$ value representing the mode confinement can be extracted from a Gaussian fit of the pulse. Mode confinement for all four simulated core diameters is shown in the inset of Fig. 1(b). The behavior is consistent with that prediction from Marcuse's equation [18] showing that smaller core diameters will experience the poorest mode confinement; see Eqn (1). Note that these values are integrated over the pulse spectrum.

Attenuation for a single frequency (1 THz) in the 140 μm core fiber was obtained by tracking the simulated flux through planes at several $z$ distances, as shown in Fig. 3(b). Intensity was integrated over the fiber core and compared with the attenuation for pure silicon. The core intensity is attenuated more strongly than would occur in silicon because $n_{mode}$ has a significant contribution from the cladding refractive index at 1 THz; see Fig. 1(c). Intensity integrated over the entire fiber was also compared with the attenuation for pure silica. Attenuation through the fiber cladding is fit well with Beer's law for pure silica.

## 5. Conclusion

The transmission of a broadband THz pulse through silica-clad, silicon-core optical fibers of various core diameters has been simulated in order to demonstrate the feasibility of this potentially practical approach. It has been determined that optimal fiber core diameters will fall approximately in the 140-250 μm range; ideally suited for the molten-core fabrication approach to semiconductor optical fibers. For smaller diameters, the guided modes will experience more spatial overlap with the cladding, especially for the lower frequency components. For larger diameters, the wavelength range for which the fibers are single-mode is narrower, requiring intermodal dispersion to be taken into account.

FDTD simulations agree well with a quantitative Beer's law absorption analysis, and with normalized frequency calculations and Marcuse's mode-confinement equation. The simulations demonstrate that THz waveguiding is achievable in semiconductor-core optical fibers of the stated core diameters.

Overall silicon-core optical fibers that can be created using a molten-core pulling technique are suitable for narrowband CW or possibly broadband pulsed THz radiation in the range of 0.5 – 3 THz. In practice the fiber-manufacturing method can be altered to use a wider range of materials beyond silicon and silica to improving THz waveguiding. Since the mode and basic propagation analysis are discussed in this proposal, the experimental proof-of-concept will be presented in detail elsewhere.

## Acknowledgements

The authors wish to thank Robert R. Rice for useful discussions. DAB and SKC were supported by a Research Challenge Grant from the West Virginia Higher Education Policy Commission (HEPC.dsr.12.29) and NSF Graduate Research Fellowship (#1102689), respectively.